\documentclass[12pt]{article}

\usepackage{graphicx}
\usepackage{epsfig}
\usepackage{natbib}
\usepackage{url}
\usepackage[shortlabels]{enumitem}                              
\usepackage{bbm}          
\usepackage{amsfonts} 
\usepackage{multirow}
\usepackage{authblk}\usepackage[margin=2.5cm]{geometry}

\usepackage{bm} 
\usepackage{dsfont} 
\usepackage{amsmath} 
\usepackage{amssymb} 
\usepackage{xifthen} 

\newcommand{\rv}[1]{\MakeUppercase{#1}} 
\newcommand{\rve}[1]{\bm{\MakeUppercase{#1}}} 
\newcommand{\stp}[2][]{\left\lbrace#2\ifthenelse{\isempty{#1}}{}{:{#1}}\right\rbrace} 

\newcommand{\ve}[1]{\bm{#1}} 
\newcommand{\m}[1]{\bm{\MakeUppercase{#1}}} 

\newcommand{\tr}{\intercal}

\newcommand{\cross}[2][]{#2^\tr\ifthenelse{\isempty{#1}}{}{#1} #2}
\newcommand{\tcross}[2][]{#2\ifthenelse{\isempty{#1}}{}{#1} #2^\tr}
\newcommand{\ccdot}{\,\cdot\,} 

\newcommand{\ind}[1]{\mathds{1}_{\left(#1\right)}} 
\newcommand{\indd}[1]{\mathds{1}{\left\lbrace#1\right\rbrace}} 

\newcommand{\mean}[2][]{\mathds{E}\ifthenelse{\isempty{#1}}{}{_{#1}}\left[#2\right]}
\newcommand{\meanhat}[2][]{\hat{\mathds{E}}\ifthenelse{\isempty{#1}}{}{_{#1}}\left[#2\right]}
\newcommand{\var}[2][]{\mathds{V}\ifthenelse{\isempty{#1}}{}{_{#1}}\left[#2\right]}
\newcommand{\cov}[3][]{\text{Cov}\ifthenelse{\isempty{#1}}{}{_{#1}}\left[#2,#3\right]}
\newcommand{\cor}[3][]{\text{Cor}\ifthenelse{\isempty{#1}}{}{_{#1}}\left[#2,#3\right]}
\newcommand{\pr}[1]{\text{Pr}\left(#1\right)} 

\newcommand{\df}[1]{\mathcal{#1}} 

\newcommand{\eq}{\text{Equation }} 
\newcommand{\eqs}{Equations} 
\newcommand{\fig}{\text{Figure }} 
\newcommand{\code}[1]{\texttt{#1}}

\title{A Model-Based General Alternative to the Standardised Precipitation Index}

\author[1]{\small Erick A. Chac\'{o}n-Montalv\'{a}n}
\author[2,3]{\small Luke Parry}
\author[2]{\small Gemma Davies}
\author[1]{\small Benjamin M. Taylor}

\affil[1]{\footnotesize Centre for Health Information, Computation and Statistics (CHICAS), Lancaster Medical School, Lancaster University, United Kingdom.}
\affil[2]{\footnotesize Lancaster Environment Centre, Lancaster University, United Kingdom.}
\affil[3]{\footnotesize Nucleus of Advanced Amazonian Studies (NAEA), Federal University of Par\'{a}, Brazil.}

\begin{document}

\maketitle

\def \mbsi {MBSI}
\def \limit {Lim}

\begin{abstract}
  In this paper, we introduce two new model-based versions of the widely-used standardized precipitation index (SPI) for detecting and quantifying the magnitude of extreme hydro-climatic events. Our analytical approach is based on generalized additive models for location, scale and shape (GAMLSS), which helps as to overcome some limitations of the SPI. We compare our model-based standardised indices (\mbsi{}s) with the SPI using precipitation data collected between January $2004$ - December $2013$ ($522$ weeks) in Caapiranga, a road-less municipality of Amazonas State. As a result, it is shown that the \mbsi{}-1 is an index with similar properties to the SPI, but with improved methodology. In comparison to the SPI, our \mbsi-1 index allows for the use of different zero-augmented distributions, it works with more flexible time-scales, can be applied to shorter records of data and also takes into account temporal dependencies in known seasonal behaviours. Our approach is implemented in an R package, \texttt{mbsi}, available from Github.

\end{abstract}

\section{Introduction}
\label{mbsi_sec:introduction}

Mitigating the effects of climate change on health and disease is one of the greatest challenges to public health and international development \citep{McMichael2013, Watts2015}. One of the main characteristics of the burden of climate change is the expected increase in the frequency, intensity and duration of extreme climate events \citep{HoughtonJT2001, Rosenzweig2001}. These are events experiencing extreme values of meteorological variables, they often cause damage and are defined as either taking maximum values or exceeding established high thresholds \citep{Stephenson2008}. In this paper, our focus is on floods and droughts, which are considered extreme hydro-climatic events because they are related to the tails of streamflow distribution \citep{Shelton2009}.

The impact of extreme hydro-climatic events is not straightforward to understand because they comprise a complex web of direct and indirect impacts on environmental, economic and social areas \citep{Blanka2017}. Floods and droughts, depending on their severity, can produce not only crucial damage to the economy and ecology of a region, but also lives can be endangered \citep{Lehner2006}. Agriculture and associated sectors are highly dependent on surface and ground water; hence, it is common to see major impacts of droughts and floods on these areas \citep{Blanka2017}. The impact of extreme hydro-climatic events will also crucially depend on specific characteristics of the society affected like their vulnerability, adaptive capacity and resilience \citep{Seiler2002, WMO2016}. Hence, focus should be put on vulnerable societies that are prone to experience extreme hydro-climatic events; for example, on roadless urban centers of the Brazilian Amazonia, where the population is experiencing droughts and floods without precedent \citep[see][]{Zeng2008, Chen2010, Filizola2014, Lewis2011}.

In this context, the importance of being able to identify extreme hydro-climatic events is due to two main reasons. First, it can help to improve the understanding of the effects of floods and droughts by allowing the analysis of extreme hydro-climatic events with respect to different variables or indicators of interest in health, economy or others.  For example, \citet{chacon2018} evaluates the effects of these events on newborn health measured through birthweight. Second, the methodology for the identification of extreme events can help to improve monitoring and prediction tools and, potentially, enhancing prevention policies to reduce the impacts of floods and droughts.

There are a large number of indices and indicators for monitoring droughts. \citet{WMO2016} presented 49 indicators and indices classified among the categories meteorology, soil moisture, hydrology, remote sensing, and composite or modelled. Between these indices, the most common are the standardized precipitation index (SPI), the Palmer drought severity index (PDSI), the crop moisture index, the surface water supply index and the vegetation condition index \citep{Mishra2010}. Comparisons between these indices, often, agree that the standardized precipitation index is an appealing index for monitoring droughts because of its simplicity, spatial invariance, probabilistic nature and its flexibility to work with different time-scales \citep{Guttman1999, Hayes1999, Morid2006, Mishra2010}. In addition, the World Meteorological Organization has suggested to use the SPI as a primary meteorological drought index through \citet{Hayes2011} and a user guide for this index has been released in \citet{WorldMetereologicalOrganization2012}.

In the case of flood monitoring, most of studies focusses on more than one indicator given that it is not only related with rainfall, but also with river levels, river discharge and geomorphology. In comparison with the case of droughts, there is not much consensus in which indices or information to use for monitoring floods. \citet{Koriche2016}, for example, used rainfall and topography to propose a satellite based index, while \citet{Ban2017} used satellite-based RGB composite imagery. Other approaches applied sensor networks or information from hydrological stations \citep{Keoduangsine2012}. Despite this variability of methodologies,  several studies are recognizing the potential value of the SPI as a tool for flood monitoring. For instance, \citet{Wang2015} demonstrated that the 2-month SPI is an effective indicator for identifying  major floods events in the Minjiang River basin. Similarly, \citet{Seiler2002, Guerreiro2008, Du2013, Koriche2016} have used the SPI for flood predicting systems.

Motivated by the desire to evaluate the impacts of extreme hydro-climatic events on birthweight in the Brazilian Amazon, \citep[see][]{chacon2018}, our research initially explored the use of the widely applied standardized precipitation index (SPI). However, although this index has been suggested as the primary meteorological drought index by the World Meteorological Organization and has been shown to be useful for identifying and monitoring droughts and floods, the current methodology for computing it has certain limitations that will be explained in Section \ref{mbsi_sub:spi_limitations}. For instance, the SPI can not be computed reliably for series shorter than 30 years. For this reason, we propose two model-based approaches that maintain the desirable characteristics of the SPI but with improved computation and methodology.

Our \emph{model-based standardized indices} (herein, \mbsi{}s) overcomes some of the limitations of the SPI by using generalized additive models for location, scale and shape (GAMLSS). These models are flexible enough to capture the seasonal trend on the parameters of the distribution of rainfall or precipitation data. Our methodology differs form other attempts to improve the SPI by proposing a model-based approach instead of proposing a group of empirical steps to compute the SPI such as presented in \citet{Erhardt2017}. In addition, the \mbsi{}s could be applied to other environmental variables of interest, other than precipitation, by choosing an appropriate family of distributions.


This paper is structured as follows. An introduction explaining the motivation for an alternative to the SPI is given in the present section. Then, the definition and limitations of the SPI are presented in Section \ref{mbsi_sec:spi}. In Section \ref{mbsi_sec:gamlss}, we provide a short introduction to generalized additive models for location, scale and shape (GAMLSS). In Section \ref{mbsi_sec:model_based_standardized_index}, two model-based approaches to compute the standardized precipitation index are proposed to tackle some of these limitations and make it possible the use of a theoretically similar index on our study for birthweight \citep[see][]{chacon2018}. After presenting the \mbsi{}s, in Section \ref{mbsi_sec:comparison}, we compare the SPI and \mbsi{}s using precipitation data collected between January 2004 - December 2013 in Caapiranga, a road-less municipality in the Amazonas State. Finally, conclusions and a discussion of the performance of our method is given in Section \ref{mbsi_sec:discussion_spi}.

\section{Standardised Precipitation Index}
\label{mbsi_sec:spi}

The SPI is an index that was proposed by \citet{Mckee1993} to improve drought detection and monitoring capabilities using statistical concepts. The main characteristics of this index is simplicity, spatial invariance, probabilistic nature and flexibility to work with different time-scales \citep{Guttman1999}. This last characteristic allows monitoring of different types of droughts like agricultural (short time-scale) and hydrological (long time-scale) \citep{Mckee1993}.

Therefore, to compute the SPI, it is necessary to choose a time-scale over which to smooth the original precipitation data; this smoothing enables the method to detect extreme events that occur over a period of continuous time. The computation of the SPI continues by evaluating the cumulative distribution function for a particular value of the smoothed precipitation, taking in consideration the seasonality, and mapping this probability to the corresponding quantile of a standard normal distribution; which is the main idea behind the SPI to quantify how extreme are the observed values with respect to the usual seasonal behaviour. The resulting series of values are interpretable in the usual manner: as quantiles from a standard normal distribution. For example, an SPI value of 2 indicates that the probability of observing an event at least as extreme as this is 0.0228. In the next sections, we describe the computation of the SPI with further detail (Section \ref{mbsi_sub:spi_methodology}), present the approach to monitor floods and droughts using the SPI (Section \ref{mbsi_sub:floods_droughts_monitoring}), and discuss some limitations of the SPI (Section \ref{mbsi_sub:spi_limitations}).

\subsection{Definition of the SPI}
\label{mbsi_sub:spi_methodology}

In this section, we outline the methodology of \citet{Mckee1993} for computing the SPI for a monthly time series of precipitation, represented as a discrete-time stochastic process, $\stp[t = 1, \dots, T]{Z_t}$. Throughout this section we will refer to $\stp{Z_t}$ as the `monthly precipitation', but the reader should bare in mind that we intend $\{Z_t\}$ to be thought of in more general terms because the methodology can, in theory, be easily applied to other variables such as river levels, river discharge, etc.

We begin by defining $\stp[t = 1, \dots, T]{X_t^k}$ as the $k$-order moving average process of $\stp{Z_t}$ such as
\begin{equation}
  \label{mbsi_eq:prec_ma}
  \rv{X}_t^k = \frac{1}{k}\sum_{i=0}^{k-1} \rv{Z}_{t-i}, ~~ \text{for} ~ t = 1, \dots, T,
\end{equation}
i.e. $x_t^k$ is the average of the observed precipitation of the last $k$ months, inclusive of the present month $t$. In the literature of drought indices, $k$ is referred to as the `time-scale' under study. The ability to define $k$ prior to analysis is considered one of the appealing characteristics of the SPI \citep{Guttman1998}.

Rather than employing formal statistical methods for selecting $k$, the choice of $k$ is determined by the time-scale under consideration by the researcher. For example, if one is interested in detecting droughts that occur over long periods of time (e.g. during a year), then $k=12$ might be chosen; similarly for analysing quarterly droughts $k=3$ might be more appropriate. The choice of time-scale can be related to the particular type of drought impact of interest. Different values of $k$ shift the focus of an analysis to different types of extreme events; this is important given that the lack of water in the short, medium or long-term affects different sections of human society and the surrounding ecosystem in different ways (e.g agricultural or hydrological effects) \citep{Mckee1993}. In the interest of disaster prevention, or planning a humanitarian response to a drought, the actions taken will be different for droughts at different time scales. For instance, events occurring on a short time-scale may be important to agricultural decisions whereas events on longer time-scales may be of more relevance for the management of water supplies \citep{Guttman1998, Guttman1999}.

To continue with the definition of the SPI, it is beneficial to switch notation for the subscript $t$, replacing $Z_t$ and $X_t^k$ by respectively $Z_{ij}$ and $X_{ij}^k$, where $i = 1, 2, \ldots, n$ is the year and $j = 1, 2, \ldots, 12$ is the month under study. We next introduce a statistical model for $X_{ij}^k$, i.e. a parametric density function, $h_j(\rv{X}_{ij}^k = x;\ccdot)$, where $x$ is an arbitrary value on the domain of $X_{ij}^k$. Notice that the notation $h_j(\ccdot;\ccdot)$ implies that the characteristics of the density function change according to the month of the year, i.e. it has a seasonal behaviour. In the original article, \citet{Mckee1993} suggested a gamma density for $h_j(\ccdot;\ccdot)$, but current practice instead makes use of a mixture, a zero-augmented gamma density (ZAGA), which allows $X_{ij}^k$ take zero values \citep{Lloyd-Hughes2002}.

Define $\pi_j=\pr{X_{ij}^k=0}$, the probability that the smoothed precipitation is zero on the month $j$, and let the density function of $X_{ij}^k$ for $X_{ij}^k>0$ be $g(\rv{X}_{ij}^k = x;\ve{\theta}_j)$, a gamma density with parameters $\ve{\theta}_j = (\mu_j, \sigma_j)^\tr$ evaluated at $x$. Thus the density function of the moving average process $X_{ij}^{k}$ is a zero-augmented gamma density defined as
\begin{equation}
  \label{mbsi_eq:ma_prec_df}
  h_j(\rv{x}_{ij}^k = x; \pi_j,\ve{\theta}_j) = \pi_j \ind{x=0} + (1-\pi_j)g(X = x;\ve{\theta}_j)\ind{x>0},
\end{equation}
where $\ind{.}$ is an indicator function. Hence, the cumulative distribution function of $X_{ij}^{k}$ is
\begin{equation}
  \label{mbsi_eq:ma_prec_distribution}
  \pr{X_{ij}^{k} \leq x} = \df{H}_j(x; \pi_j, \ve{\theta}_j) =
  \left\{
  \begin{array}{ll}
  \pi_j & x = 0\\
  \pi_j + (1-\pi_j)\df{G}(x; \ve{\theta}_j) & x > 0
  \end{array}
  \right.,
\end{equation}
where $\df{G}(\ccdot; \ve{\theta}_j)$ denotes the distribution function for a gamma random variable with parameters $\ve{\theta}_j$.

A key point we will revisit in the sequel is that the parameters $\pi_j$ and $\ve{\theta}_j$ in \eqs{} \ref{mbsi_eq:ma_prec_df} and \ref{mbsi_eq:ma_prec_distribution} vary from month to month, but not between years, so they are able to capture annual seasonal behaviours. The methodology of \citet{Mckee1993} thus partitions $\stp{X_{ij}^t}$ into twelve independent series of the form $\rve{X}_{[j]}^k = (\rv{X}_{1j}^k, \rv{X}_{2j}^k, \dots, \rv{X}_{nj}^k)^\tr$ for $j = 1, \dots, 12$. Parameter estimation for each month, $\hat{\pi}_j$ and $\hat{\ve{\theta}}_j$, is done independently by fitting a realisation of $\rve{X}_{[j]}^k$, i.e. $\ve{x}_{[j]}^k = (x_{1j}^k, x_{1j}^k, \dots, x_{nj}^k)$, to the zero-augmented gamma density $h_j(\ccdot; \ccdot)$ in \eq \ref{mbsi_eq:ma_prec_df}.

Values of the standardized precipitation index (SPI) are then obtained by computing the quantiles for a standard normal density with probabilities $\df{H}_j(\cdot; \cdot)$. As mentioned before, the interpretation of obtained SPI values are as the one of a standard normal distribution, e.g. values greater than $3$ or lower than $-3$ can be considered extreme values, while values close to zero are likely to happen.

Provided $h_j(\ccdot; \ccdot)$ fits the data well, for each $j$, the probability integral transform implies we should expect the collection $\Pi=\{\df{H}_j(x_{ij}^k;\hat\pi_j,\ve{\hat\theta}_j)\}$ to follow a standard uniform density; the back-transform using the inverse cumulative distribution function of a standard Gaussian is therefore redundant, beyond relating $\Pi$ to quantiles of a density commonly used in statistical practice.


Hence, the proposed method of \citet{Mckee1993} to compute the SPI can be summarized as:
\begin{enumerate}[1), nolistsep]
  \item Define the time-scale $k$ to work with (e.g. 1 month, 3 months, etc).
  \item Compute the $k$-order moving average series $\stp{x_{ij}^k}$ using all the precipitation time series $\stp{z_{ij}^k}$.
  \item Split the moving average series $\{x_{ij}^k\}$ into months to obtain $\ve{x}_{[1]}^k$, $\ve{x}_{[2]}^k$, $\dots$, $\ve{x}_{[12]}^k$.
  \item For each month $j$, obtain the estimates $\hat{\pi}_j$ and $\hat{\ve{\theta}}_j$ by fitting the realization of $\ve{X}_{[j]}^k$, i.e. $\ve{x}_{[j]}^k$, to the density function $h_j(\ccdot; \ccdot)$ on \eq \ref{mbsi_eq:ma_prec_df}. Maximum likelihood estimation can be used for this step.
  \item Evaluate the cumulative density function $\df{H}(\ccdot; \ccdot)$ to the observed values of the moving average process $\{X_{ij}^k\}$ to obtain the collection $\Pi=\{\df{H}_j(x_{ij}^k;\hat\pi_j,\ve{\hat\theta}_j)\}$.
  \item Obtain the values for the SPI by computing the quantiles of a standard normal distribution with probabilities $\Pi=\{\df{H}_j(x_{ij}^k;\hat\pi_j,\ve{\hat\theta}_j)\}$.
\end{enumerate}

\subsection{Flood and Drought Monitoring}%
\label{mbsi_sub:floods_droughts_monitoring}

For drought monitoring, \citet{Mckee1993} defined an episode of \textit{drought} as a period of time in which the SPI is continuously negative reaching at least one value lower or equal to $-1$. Then, it is said that the beginning of the drought is the first time that the SPI falls below zero and it finishes when a positive SPI is reached after observing a value lower or equal to $1$ \citep{Mckee1993}. Similarly, a \textit{flood} can be defined as a period of time where the SPI is continuously positive reaching at least one value greater or equal to $1$. Further characteristics of these events, such as \emph{magnitude} and \emph{intensity}, can be computed to improve drought monitoring. For example, the magnitude has been defined as the absolute value of the sum of the SPI during the period of the drought/flood, while the intensity can be classify as shown in table \ref{mbsi_tab:spi_table} \citep{Mckee1993, Wang2015}.
\begin{table}[!ht]
  \centering
  \caption{Intensity of droughts and floods based on the SPI}
  \label{mbsi_tab:spi_table}
  \begin{tabular}{|l|c|}
    \hline
    \textbf{Category} & \textbf{Value} \\ \hline
    \hline
    extreme flood & $ \text{SPI} \geq2$ \\ \hline
    severe flood & $1.5\leq \text{SPI} <2$ \\ \hline
    moderate flood & $1\leq \text{SPI}<1.5$ \\ \hline
    near normal & $-1<\text{SPI}<1$ \\ \hline
    moderate drought & $-1.5<\text{SPI} \leq-1$ \\ \hline
    severe drought & $-2<\text{SPI} \leq-1.5$ \\ \hline
    extreme drought & $\text{SPI} \leq2$ \\ \hline
  \end{tabular}
\end{table}

\subsection{Limitations of the SPI}
\label{mbsi_sub:spi_limitations}

The standardised precipitation index has the following main limitations (\limit{}):

\begin{enumerate}[(\limit{} 1), ref = \limit{} \arabic*, nolistsep, leftmargin=*]
  \item \label{mbsi:lim_gamma} \emph{The zero-augmented gamma distribution might not be a good fit for the precipitation data:} Although in most practical applications the zero-augmented gamma distribution has been observed to be a good choice for precipitation data, there have been cases where it has been found to be inadequate \citep{Guttman1999, Mishra2010}. While it might be straightforward in theory to extend the standard SPI model to include other distributional choices for $h(\cdot; \cdot)$, it would nevertheless be useful if the methodology itself was more flexible in this regard.
  \item \label{mbsi:lim_scale} \emph{The time-scale is based on months:} Theoretically there is no impediment to work with a time-scale other than months, but most published studies do not do this. Additionally, the official SPI user guide recommends working with a time-scale of at least 4 weeks (1 month) stating that lower values will make the SPI behave more erratically \citep{WorldMetereologicalOrganization2012}. It would be desirable to develop an index that is flexible enough to allow the use of shorter and more arbitrary time-scales.
  \item \label{mbsi:lim_long} \emph{It requires a long record of precipitation:} In order to compute the SPI, it is recommended that at least 30 years of precipitation records are available, and ideally between 50 and 60 years \citep{PiratheeparajahN2014}. The reason for this is the splitting of the complete moving average series into 12 independent subsets corresponding to each month of the year. Each of these twelve subsets has length equal to the number of years $n$ under study, therefore small values of $n$ may not provide reliable estimates of $\pi_j$ and $\ve{\theta}_j$. This problem is related to the fact that subsets of data are handled independently.
  \item \label{mbsi:lim_correlation} \emph{It ignores the temporal correlation and the cyclic nature of $Z_t$, and hence in $X_t^k$, (i.e. we would expect $X_{i,12}^k$ to be correlated with $X_{i+1,1}^k$):} It is natural to observe a correlated and cyclic behaviour on precipitation data and the parameters associated with the density function; however, the SPI does not take this into account. This could affect parameter estimation because an outlier presented in certain month could drastically affect the estimated value of the parameters for that month only. This way the parameters will not vary smoothly across neighbouring months, which is both an undesirable property, but also affects the reliability of SPI values. In neglecting the temporal correlation inherent in time series such as precipitation, the SPI does not take advantage of the fact that time is a continuous variable and the the continuous sharing of information across time should improve parameter estimation and allow us to work with shorter time series (which is related to \ref{mbsi:lim_long}).
\end{enumerate}

\section{Generalized Additive Models for Location, Scale and Shape}
\label{mbsi_sec:gamlss}

In this paper we suggest the use of generalized additive models for location, scale and shape (GAMLSS) to tackle the limitations of the SPI presented in Section \ref{mbsi_sub:spi_limitations}. We briefly introduce this type of model in the present section.

A generalized additive model (GAM) is an extension of an generalized linear model (GLM) that allows for the inclusion of smooth functions of covariates into the linear predictor \citep{Rigby2005,Umlauf2017} and thus they allow complex relationships between predictors and outcomes to be captured. The smooth functions are defined as linear combinations of basis functions, the most common being cubic regression splines, P-splines, thin plate regression splines and tensor product splines \citep{Wood2006}.

A GAMLSS is an extension of a GAM where, in addition to the location parameter, the scale and shape parameter are also modeled with respect to covariates. More formally, assuming a response variable $\rv{Y}_i$ with probability density function $f(y_i|\theta_{i1}, \dots, \theta_{iK})$, each parameter $\theta_{ik}$ for $k = 1, \dots, K$ is associated to a linear predictor $\eta_{ik}$ through a monotonic link function $g_k$ such as
\begin{align}
  \label{mbsi_eq:linear_predictor_gamlss_individual}
  \begin{split}
    g_k(\theta_{ik}) = \eta_{ik} =
    \ve{x}_{i0k}^\intercal\ve{\beta}_{0k} +
    f_{1k}(\m{x}_{i1k}; \ve{\beta}_{1k}) +
    \dots +
    f_{J_kk}(\m{x}_{iJ_kk}; \ve{\beta}_{J_kk}),
  \end{split}
\end{align}
where $\ve{\beta}_{0k}$ represents the fixed effects associated to the covariates $\ve{x}_{i0k}$ for an individual $i$, and $f_{jk}$ represent functions able to capture a wide variety of effects with corresponding parameters $\ve{\beta}_{jk}$ and covariates $\ve{x}_{ijk}$. Note that $f_{jk}(\ve{x}_{ijk}; \ve{\beta}_{jk})$ can not only represent smooth functions of a covariate, but also smooth functions of multiple covariates or varying effects. Considering $h(.)$ as a smooth function, $f_{jk}$ can be used to represent: a smooth effect $h(x)$, varying coefficient $x_1\times h(x_2)$, a smooth multiple effect $h(x_1, \dots, x_L)$, a random intercept $b_g$, a random slope $x\times b_g$, a spatial effect $h(\code{lat}, \code{lon})$, a temporal effect $h(\code{time})$, a space-time effect $h(\code{lat}, \code{long}, \code{time})$, and others such as seasonal effects \citep{Umlauf2017}.

More generally, for a set of observations $y_1, \dots, y_n$, parameter vector $\ve{\theta}_k = (\theta_{1k}, \dots, \theta_{nk})$ and linear predictor vector $\ve{\eta}_k = (\eta_{1k}, \dots, \eta_{nk})$, we can rewrite Equation \ref{mbsi_eq:linear_predictor_gamlss_individual} in matrix form as
\begin{align}
  \label{mbsi_eq:linear_predictor_gamlss}
  \begin{split}
    g_k(\ve{\theta}_{k}) = \ve{\eta}_{k} =
    \m{X}_{0k}\ve{\beta}_{0k} +
    f_{1k}(\m{X}_{1k}; \ve{\beta}_{1k}) +
    \dots +
    f_{J_kk}(\m{X}_{J_kk}; \ve{\beta}_{J_kk}),
  \end{split}
\end{align}
such as $\m{X}_{0k}$ represents the design matrix with fixed effects $\ve{\beta}_{0k}$ and $\m{X}_{jk}$ is the design matrix required to construct the effect $f_{jk}$ with parameters $\ve{\beta}_{jk}$. The structure of $\m{X}_{jk}$ will depend on the type of effects that are desired to be captured by $f_{jk}$ as well as the type of covariates involved. Although \citet{Umlauf2017} allows $f_{jk}(\ve{x}_{ijk}; \ve{\beta}_{jk})$ to take more complex structures, the most common type of effects and the ones included in this study take the form $f_{jk}(\ve{x}_{ijk}; \ve{\beta}_{jk}) = \m{X}_{jk} \ve{\beta}_{jk}$.

Estimation usually proceeds using a penalised likelihood approach \citep{Rigby2005,Wood2006}, or a Bayesian approach \citep{Umlauf2017}.

\section{A Model-Based Method for Evaluating Extreme Hydro-Climatic Events}
\label{mbsi_sec:model_based_standardized_index}

%

Having discussed some of the shortcomings of the SPI, in this section we propose two alternatives to the SPI: these will be model-based approaches which we refer to as the \emph{model-based standardised indices} (\mbsi s: \mbsi-1 and \mbsi-2); we argue that our indices retain the desirable characteristics of the SPI, but improve the methodology. Our model-based standardised indices are more stable, flexible and satisfying (from a modelling perspective) than the SPI as explained in Section \ref{mbsi_sec:discussion_spi}.

Although there have been attempts to improve the methodology of the SPI (e.g. \citet{Erhardt2017, WMO2016}) our method differs because we use a model-based approach, which accounts for the characteristics required to compute a standardized index. In contrast, \citet{Erhardt2017} proposed a group of steps to compute the SPI including;  elimination of seasonality (including variable transformation to reduce skewness, computation of monthly sample and variance mean) and elimination of temporal dependence and transformation to the standard normal distribution. Our model-based approach allows us to not only compute the standardized precipitation index appropriately, but also enables us to work with short time series, check assumptions, work at different scales (e.g. weeks), work with missing values and provide further relevant information about the underlying process under study, such as precipitation.

In Section \ref{mbsi_sub:model_based_standardized_index_i} and \ref{mbsi_sub:model_based_standardized_index_ii} respectively, we introduce the model-based standardised index (\mbsi-1) and (\mbsi-2), which address the limitations of the SPI discussed above using generalized additive models for location, scale and shape (GAMLSS). In addition, we discuss some limitations of using GAMLSS in Section \ref{mbsi_sub:limitations_gamlss}.

\subsection{Model-based Standardized Index 1 (\mbsi-1)}
\label{mbsi_sub:model_based_standardized_index_i}

In Section \ref{mbsi_sub:spi_methodology}, we saw that the SPI is defined for the moving average process $\stp{X_{ij}^{k}}$ of a discrete stochastic process $\stp{Z_{ij}}$, where $i$ denoted the year and $j$ the month. The \mbsi-1 instead uses the initial notation of Equation \ref{mbsi_eq:prec_ma}, i.e. we work directly with $\stp[t=1,\dots,T]{Z_t}$ and $\stp[t=1,\dots,T]{X_{t}^{k}}$ as the precipitation and moving average process respectively. Note that we are assuming that $t$ and $k$ are on the same scale, which can be an arbitrary one such as daily, weekly, monthly, etc.

For the \mbsi-1, we again define the density function of each element of the stochastic process $\stp{X_{t}^{k}}$ as a mixture such as
\begin{equation}
  \label{mbsi_eq:ma_prec_df2}
  h(\rv{x}_t^k = x; \pi_t,\ve{\theta}_t) = \pi_t \ind{x=0} + (1-\pi_t)g(\rv{X}_t^k = x;\ve{\theta}_t)\ind{x>0},
\end{equation}
where $x$ is an arbitrary value on the domain of $X_t^k$, while $\pi_t$ and $\ve{\theta}_t$ are the parameters associated with the mixture density at time $t$.

The density function $g(\ccdot;\ccdot)$ can be any distribution defined on the positive real numbers that is adequate for characterizing the moving average precipitation. In this paper, we complete the definition of $h$ by using a gamma density for $g(\ccdot;\ve{\theta}_t)$ with parameters $\ve{\theta}_t = (\mu_t, \sigma_t)^\tr$, defined as follows
\begin{equation}
  \label{mbsi_eq:gamma}
g(x_t^k; \mu_t,\sigma_t) = \frac{(\sigma_t/\mu_t)^{\sigma_t}}{\Gamma(\sigma_t)} x^{\sigma_t-1} \exp\left(-\frac{\sigma_t}{\mu_t}x\right).
\end{equation}
However, note that our approach is not limited to this distribution, and a different choice of $g(\cdot;\ve{\theta}_t)$ may be more suitable in other situations. One consequence of assuming a gamma density is that the mean and variance of $[X_{t}^k|X_{t}^k>0]$ are $\mu_t$ and $\mu_t^2/\sigma_t$ respectively.

As mentioned earlier, the SPI tries to quantify the extremity of levels of precipitation by comparing it with the usual seasonal behaviour. For this reason, our approach captures the seasonal behaviour in all the parameters by introducing models for $\pi_t$, $\mu_t$ and $\sigma_t$, as in \eq \ref{mbsi_eq:linear_predictor_gamlss_individual}, using linear predictors $\eta_{1t}$, $\eta_{2t}$ and $\eta_{3t}$ such as
\begin{align}
  \label{mbsi_eq:linear_predictor_MBSI}
  \begin{split}
    \log\left(\frac{\pi_{t}}{1-\pi_{t}}\right) & =  \eta_{1t} = \m{X}_1\ve{\alpha}_1 + f_1(t; \ve{\beta}_1),\\
    \log(\mu_{t})& = \eta_{2t} = \m{X}_2\ve{\alpha}_2 + f_2(t; \ve{\beta}_2),\\
    \log(\sigma_{t})& = \eta_{3t} = \m{X}_3\ve{\alpha}_3 + f_3(t; \ve{\beta}_3),\\
  \end{split}
\end{align}
where $\m{X}_1$, $\m{X}_2$ and $\m{X}_3$ are (optional) design matrices that include information for predicting the process with linear effects $\ve{\alpha}_1$, $\ve{\alpha}_2$ and $\ve{\alpha}_3$; and $\ve{\beta}_1$, $\ve{\beta}_2$ and $\ve{\beta}_3$ are the parameters required to define the flexible non-linear functions $f_1(\ccdot;\ccdot)$, $f_2(\ccdot;\ccdot)$ and $f_3(\ccdot;\ccdot)$ that capture the seasonal effects on $\pi_t$, $\mu_t$ and $\sigma_t$ respectively. A common choice for these functions in the generalised additive modelling literature is to represent them using \textit{cyclic cubic splines} \citep{Wood2006}. An alternative to cyclic cubic splines is to use harmonic terms to represent seasonal effects. However, our experience of harmonic models in this context tends to overfit the data and using stepwise selection to reduce the number of harmonic terms can be a computationally slow process.

Our model, defined with \eqs{} \ref{mbsi_eq:ma_prec_df2}, \ref{mbsi_eq:gamma} and \ref{mbsi_eq:linear_predictor_MBSI} is a generalized additive model for location, scale and shape (GAMLSS), as explained in Section \ref{mbsi_sec:gamlss}, using a zero-augmented gamma likelihood (ZAGA). Inference can be achieved using standard methods: backfitting or MCMC \citep{Rigby2005, Umlauf2017}.

Another option for modelling serial dependence in the parameter vector $\ve{\theta}_t$ would be to assume a latent, possibly multivariate, Gaussian process or a moving average process for $f_1(\ccdot; \ccdot)$, $f_2(\ccdot; \ccdot)$ and $f_3(\ccdot; \ccdot)$. We have not explored these options, but they fit into the class of latent Gaussian models, for which there are a range of model fitting options, including INLA, MCMC and particle filtering, if not off-the-shelf software solutions to implement them.

Once we have estimated the parameters in our models, we can predict $\pi_t$, $\mu_t$ and $\sigma_t$ for any time $t$ and proceed with the computation of the \mbsi-1 using steps 5 and 6 of Section \ref{mbsi_sub:spi_methodology}. Hence, the computation of standardised precipitation values using \mbsi-1 can be summarized with the following steps:
\begin{enumerate}[1), nolistsep]
  \item Define the time-scale $k$ to work with (e.g. 1 week, 4 weeks, 8 weeks, etc).
  \item Compute the $k$-order moving average series $\{x_t^k\}$ using all the precipitation time series $\{z_{t}^k\}$.
  \item Obtain the parameters estimates $\hat{\ve{\beta}}_1$, $\hat{\ve{\beta}}_2$, $\hat{\ve{\beta}}_3$, $\hat{\ve{\alpha}}_1$, $\hat{\ve{\alpha}}_2$ and $\hat{\ve{\alpha}}_3$ by fitting the moving average series $\{x_t^k\}$ to the GAMLSS model with zero-augmented gamma distribution (\eqs{} \ref{mbsi_eq:ma_prec_df2} and \ref{mbsi_eq:gamma}) and linear predictors defined in \eq \ref{mbsi_eq:linear_predictor_MBSI}.
  \item With the parameters estimated in the previous step ($\hat{\ve{\beta}}_1$, $\hat{\ve{\beta}}_2$, $\hat{\ve{\beta}}_3$, $\hat{\ve{\alpha}}_1$, $\hat{\ve{\alpha}}_2$ and $\hat{\ve{\alpha}}_3$), obtain the estimates $\hat{\pi}_t$ and $\hat{\ve{\theta}}_t$, using \eq \ref{mbsi_eq:linear_predictor_MBSI}, for $t = 1, \dots, T$.
  \item Evaluate the cumulative density function $\df{H}(\cdot; \cdot)$ of the observed values of the moving average process $\{X_{t}^k\}$ to obtain the collection $\Pi=\{\df{H}(x_{t}^k;\hat\pi_t,\ve{\hat\theta}_t)\}$.
  \item Obtain the values for the SPI by computing the quantiles of a standard normal distribution with probabilities $\Pi$.
\end{enumerate}

\subsection{Model-based Standardized Index 2 (\mbsi-2)}
\label{mbsi_sub:model_based_standardized_index_ii}

One disadvantage of the \mbsi-1 is that it requires to fit a model to the moving average process $\{X_{t}^k\}$ for every scale-time of interest $k$. As an alternative to the \mbsi-1, we propose a second approach under which the model fitting in only done once, for $k=1$. We will refer to this approach as the model-based standardised index 2 (\mbsi-2).

For this second approach, instead of imposing a model on the elements of the moving average process $\stp{X_{t}^k}$, we propose a model for the original stochastic process $\stp{Z_t}$ that represents the precipitation. Specifically, we assume that $\rv{Z}_t$ has a zero-augmented gamma distribution, which is defined by \eqs{} \ref{mbsi_eq:ma_prec_df2} and \ref{mbsi_eq:gamma}, and the parameters are modelled considering a seasonal behaviour as in \eq \ref{mbsi_eq:linear_predictor_MBSI}. As a consequence, we can obtain the distribution of moving average $X_t^k =\sum_{i=0}Z_{t-i}/n$ for each $t \geq k$. Unfortunately, we can not obtain the analytical expression of the resulting distribution of $\rv{X}_t^k$. However, we can use Monte Carlo methods to obtain the cumulative distribution function $\df{H}(\ccdot;\ccdot)$ evaluated on the observed values of the moving average process $\stp{X^k_t}$, obtaining $\Pi=\{\df{H}(x_{t}^k;\hat\pi_t,\ve{\hat\theta}_t)\}$. Once these probabilities are obtained, we can finally obtain the quantiles of a standard normal distribution associated to these probabilities $\Pi$.

Hence, the computation of the \mbsi-2 can be summarized as follows:
\begin{enumerate}[1), nolistsep]
  \item Obtain the parameters estimates $\hat{\ve{\beta}}_1$, $\hat{\ve{\beta}}_2$, $\hat{\ve{\beta}}_3$, $\hat{\ve{\alpha}}_1$, $\hat{\ve{\alpha}}_2$ and $\hat{\ve{\alpha}}_3$ by fitting the original precipitation series $\{z_t^k\}$ to the GAMLSS model with zero-augmented gamma distribution (\eqs{} \ref{mbsi_eq:ma_prec_df2} and \ref{mbsi_eq:gamma}) and linear predictors defined in \eq \ref{mbsi_eq:linear_predictor_MBSI}.
  \item With the parameters estimated in the previous step ($\hat{\ve{\beta}}_1$, $\hat{\ve{\beta}}_2$, $\hat{\ve{\beta}}_3$, $\hat{\ve{\alpha}}_1$, $\hat{\ve{\alpha}}_2$ and $\hat{\ve{\alpha}}_3$), obtain the estimates $\hat{\pi}_t$ and $\hat{\ve{\theta}}_t$, using \eq \ref{mbsi_eq:linear_predictor_MBSI}, for $t = 1, \dots, T$.
  \item Obtain $m$ realizations $\{z_t^{(l)}\}$, where $l=1, \dots, m$,  of the precipitation stochastic process $\stp{Z_t}$ using $\hat{\pi}_t$ and $\hat{\ve{\theta}}_t$ for a zero-augmented gamma distribution (\eqs{} \ref{mbsi_eq:ma_prec_df2} and \ref{mbsi_eq:gamma}).
  \item Define the time-scale $k$ to work with (e.g. 1 week, 4 weeks, 8 weeks, etc).
  \item Compute the $k$-order moving average series $\{x_t^k\}$ of the precipitation time series $\{z_{t}^k\}$ and the $k$-order moving average series $\{x_t^{k^{(l)}}\}$ of the $m$ samples $\{z_t^{(l)}\}$.
  \item Evaluate the cumulative density function of the observed values of the moving average process $\{X_{t}^k\}$ to obtain the collection $\Pi=\{\df{H}(x_{t}^k;\hat\pi_t,\ve{\hat\theta}_t)\}$ considering that
    \begin{equation}
      \nonumber
      \df{H}(\rv{X}_t^k = x_t^k; \hat{\pi}_t, \hat{\ve{\theta}}_t) = \pr{\rv{X}_t^k \leq x_t^k} = \sum_{l=1}^m \frac{\indd{x_t^{k^{(l)}} < x_t^k}}{m}.
    \end{equation}
  \item Obtain the values for the SPI by computing the quantiles of a standard normal distribution with probabilities $\Pi$.
\end{enumerate}

\subsection{Limitations of GAMLSS}
\label{mbsi_sub:limitations_gamlss}

Although generalized additive models are attractive, they have some limitations that are worth exploring. Firstly, there can be a tendancy to overfitting the data, for example, it is known that the generalized cross-validation criterion used to estimate the smoothing parameter can lead to overfitting; however, this is less likely with a large number of observations and when the values across covariates are very well distributed \citep{Wood2006}. This problem can worsen when modelling in addition the scale and shape parameters because the model is much more flexible and appropriate precaution should be exercised on small sample sizes.

Another limitation is that prediction outside the range of values observed on the covariates might not be reliable because usually few observations with extreme values in the covariates are observed. Given that the model is very flexible, it will try to adapt to these values. In this way, prediction at the tails of the covariates may vary significantly from one sample to another, indicating that the model has high variance in the tails of covariates. Nevertheless, extrapolation is also problematic in other types of models.

Finally, the interpretability of GAM models is not as easy for GLM models and it is required to visualize the effects in order to understand and interpret them. Despite this, we view the visualization process as actually provide useful information on the effects at different levels. Also, when using credible intervals, insight into the significance of each term is obtained.

In conclusion, GAM and GAMLSS are attractive models, but they should be used with precaution given that their inherent flexibility.

\section{Comparison Between the SPI and \mbsi}%
\label{mbsi_sec:comparison}

In order to illustrate differences between the SPI, \mbsi-1 and \mbsi-2, we compare parameter estimation, model checking and the resulting standardized precipitation values for different time-scales using data collected between January $2004$ - December $2013$ (522 weeks) in Caapiranga, a road-less municipality in Amazonas State.

We use our \texttt{R} package \texttt{mbsi}, created to analyse and visualise extreme events, available from Github, \url{https://github.com/ErickChacon/mbsi}. It contains the implementation of the SPI, \mbsi-1 and \mbsi-2 indices used in this section.

\subsection{Parameter Estimation}
\label{mbsi_sub:parameter_estimation}

\begin{figure}[!ht]
  \centering
  \includegraphics[width=1\linewidth]{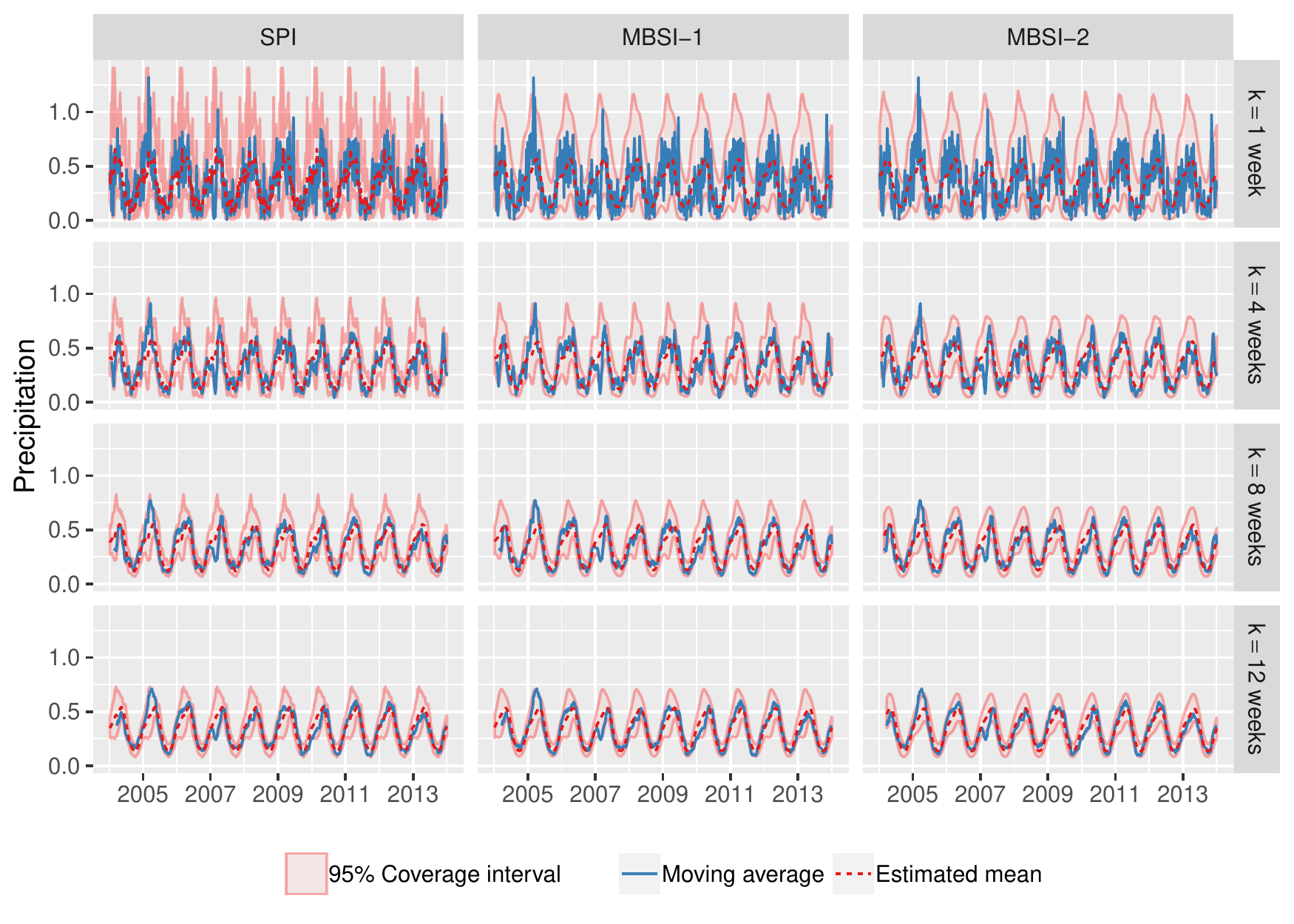}
  \caption{Precipitation moving average and $95\%$ coverage interval obtained by the SPI, \mbsi-1 and \mbsi-2 methodologies for different time-scales ($1$, $4$, $8$ and $12$ weeks)}
  \label{mbsi_fig:rain_moving_average}
\end{figure}

In this section we compare the estimated mean and coverage interval of the moving average rainfall  $X_t^k$ obtained with the estimated parameters using both the SPI and the MBSI methodologies (Fig. \ref{mbsi_fig:rain_moving_average}). Given the density function defined in Equation (\ref{mbsi_eq:ma_prec_df2}) with Gamma density $g(.;\rve{\theta}_t)$, the estimated mean for time $t$ is $(1-\hat{\pi}_t)\hat{\mu}_t$ and the $95\%$ coverage interval for a time $t$ is obtained by computing the $0.025$ and $0.975$ quantiles of the estimated density function $h(x_{ij}^k; \hat{\pi}_j, \hat{\ve{\theta}}_j)$.

We can see in \fig \ref{mbsi_fig:rain_moving_average} that at a time-scale of 1 week, the mean and coverage interval change  quickly for the classical SPI, whereas they change smoothly for the \mbsi{}s. This is an indication that the SPI overfitted the observed precipitation data. Another characteristic of the SPI at this shorter time-scale is that parameter estimation is strongly affected by extreme short-term values. The coverage interval is highly influenced by these extreme values leading sometimes to much wider coverage intervals (e.g. due to some observations around 2005). This can reduce the ability of the SPI to detect extreme events, e.g. it can be seen in Figure \ref{mbsi_fig:rain_moving_average} that there are more values lying outside the coverage intervals for the \mbsi{}s. Both characteristics happen because parameters in the SPI are independent among months, while the \mbsi{}s explicitly model this dependence using smooth functions.

As the time-scale increases, the difference between the estimated mean and coverage intervals methods decrease, but the coverage intervals are still wider and looser for the SPI.

\subsection{Model Checking}%
\label{mbsi_sub:model_checking}

\begin{figure}[!ht]
  \centering
  \includegraphics[width=1\linewidth]{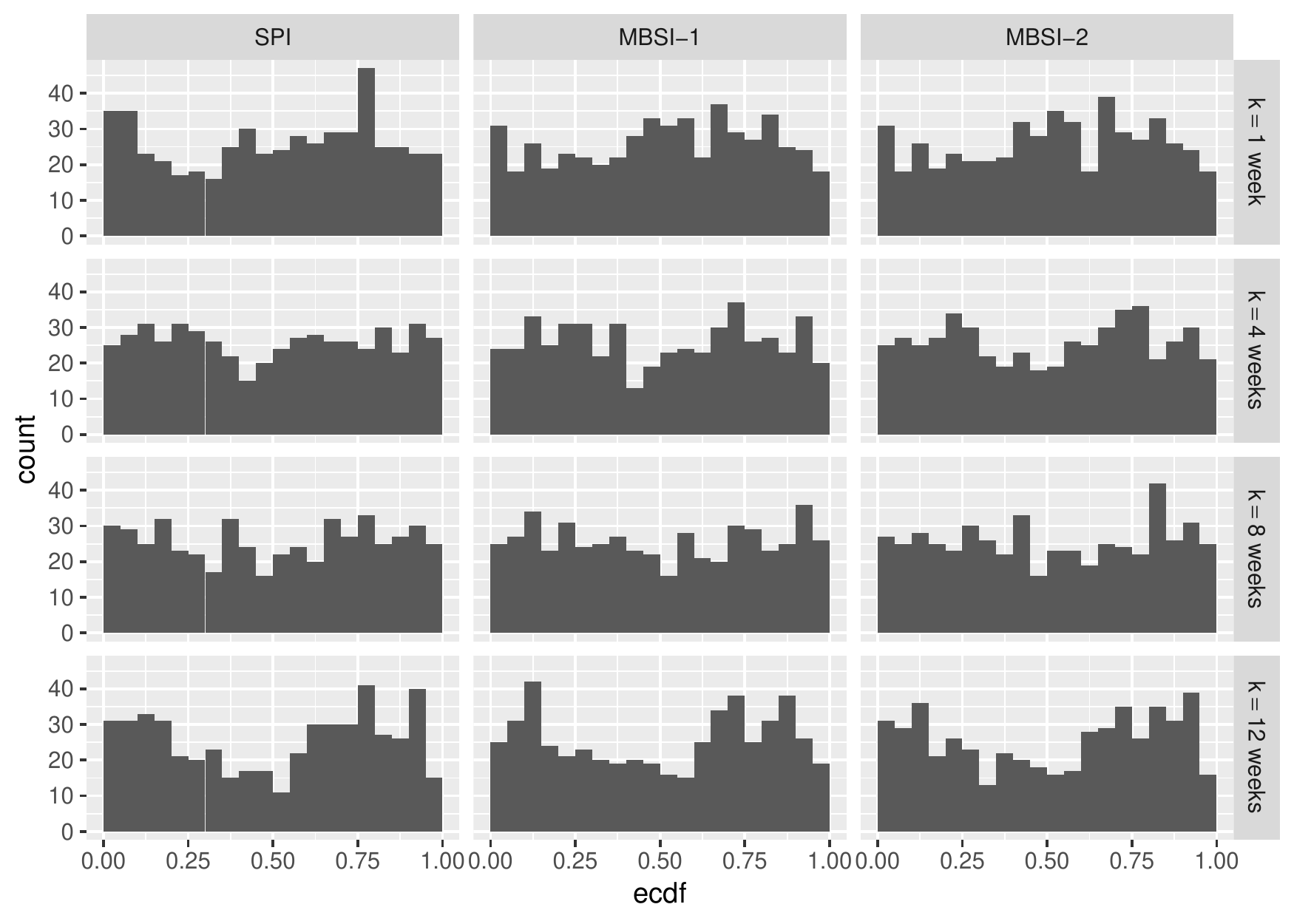}
  \caption{Distribution of the empirical cumulative density function $\Pi=\{\df{H}(x_{ij}^k;\hat\pi_j,\ve{\hat\theta}_j)\}$ for the SPI, \mbsi-1 and \mbsi-2 methodologies for different time-scales ($1$, $4$, $8$ and $12$ weeks). It should have a uniform distribution when the underlying model is adequate.}
  \label{mbsi_fig:probability_integral_transform}
\end{figure}

Provided the assumed density function $h(\ccdot; \ccdot)$ fits the data well, the probability integral transform implies we should expect the collection of the empirical cumulative density values, $\Pi=\{\df{H}(x_{ij}^k;\hat\pi_j,\ve{\hat\theta}_j)\}$, to follow a standard uniform density. If this does not hold, then the interpretation as a standard normal distribution of the standardized values is misleading since the back-transformed data will not be normally distributed. By inspecting \fig \ref{mbsi_fig:probability_integral_transform}, we can see that the uniform assumption holds for the three approaches for $k = 1, 4, 8$ weeks, while for $k=12$ weeks, it seems that the \mbsi{}s are more adequate. In general, there is no indication of drastic inadequacies for any of the methodologies.

\begin{figure}[!ht]
  \centering
  \includegraphics[width=1\linewidth]{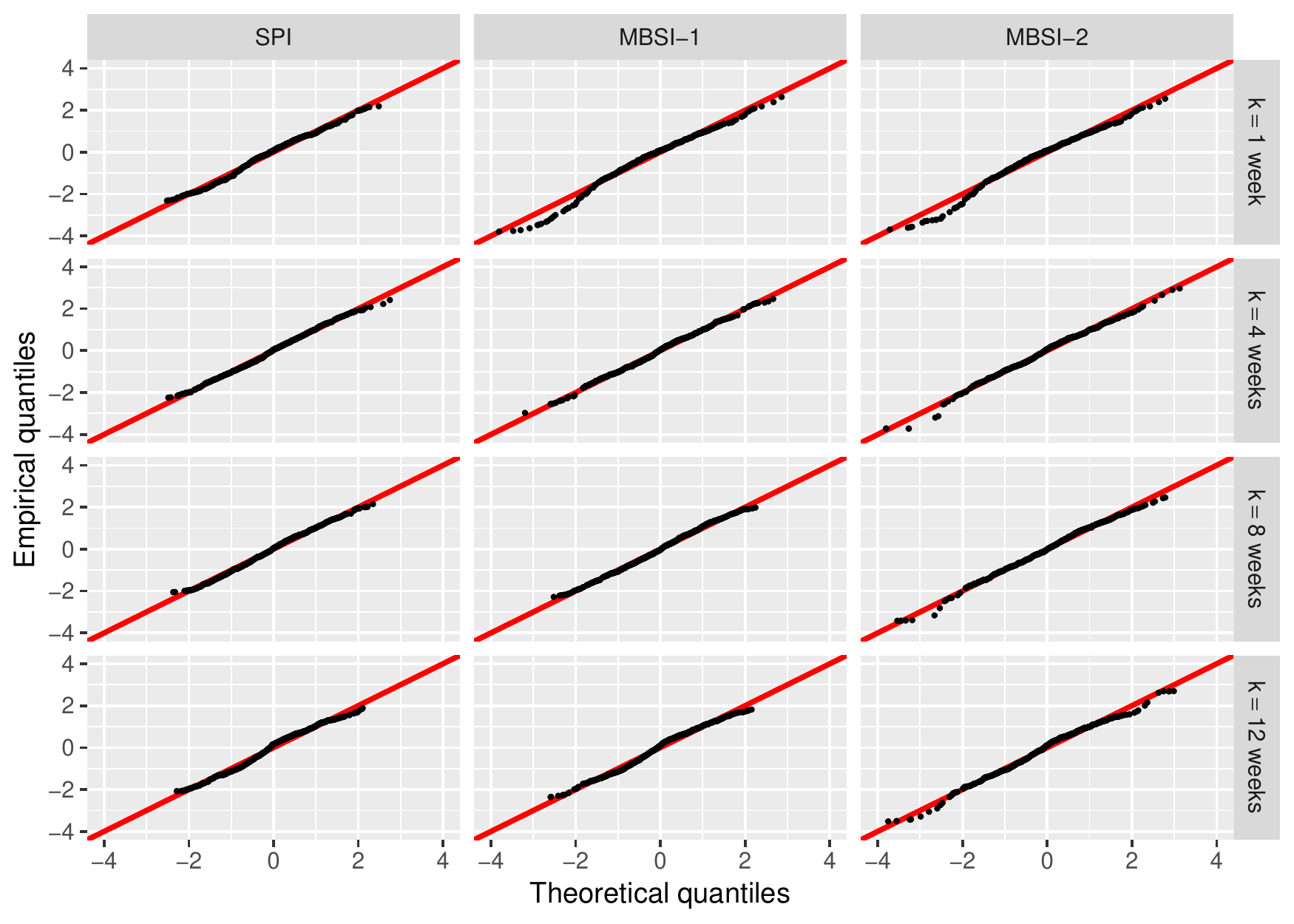}
  \caption{Comparison between the empirical quantiles (standardized precipitation values) and theoretical quantiles of a standard normal distribution for the SPI, \mbsi-1 and \mbsi-2 methodologies for different time-scales ($1$, $4$, $8$ and $12$ weeks). The points should be close to the identity line (straight line) to hold the assumption of normality.}
  \label{mbsi_fig:qqplot}
\end{figure}

If the uniformity assumption holds, then under the probability integral transform theorem, the obtained standardized precipitation values should follow a standard normal distribution, which can be checked by comparing the empirical quantiles with the theoretical quantiles of a standard normal distribution as shown in \fig \ref{mbsi_fig:qqplot}. Although, we can see in \fig \ref{mbsi_fig:qqplot} that there are some small deviations from the identity line for the \mbsi-1 and \mbsi-2 at small scales, the points lie close to the identity line for the three methodologies and the four time-scales. Something to notice is that the SPI methodology tends to limit the standardized values between $2$ and $-2$ for this data of $522$ observations, while we obtain more extreme standardized values with the \mbsi{}s, something highlighted even more for the \mbsi-2.

\subsection{Standardized Precipitation Values}%
\label{mbsi_sub:standardized_precipitation_values}

\begin{figure}[!ht]
  \centering
  \includegraphics[width=1\linewidth]{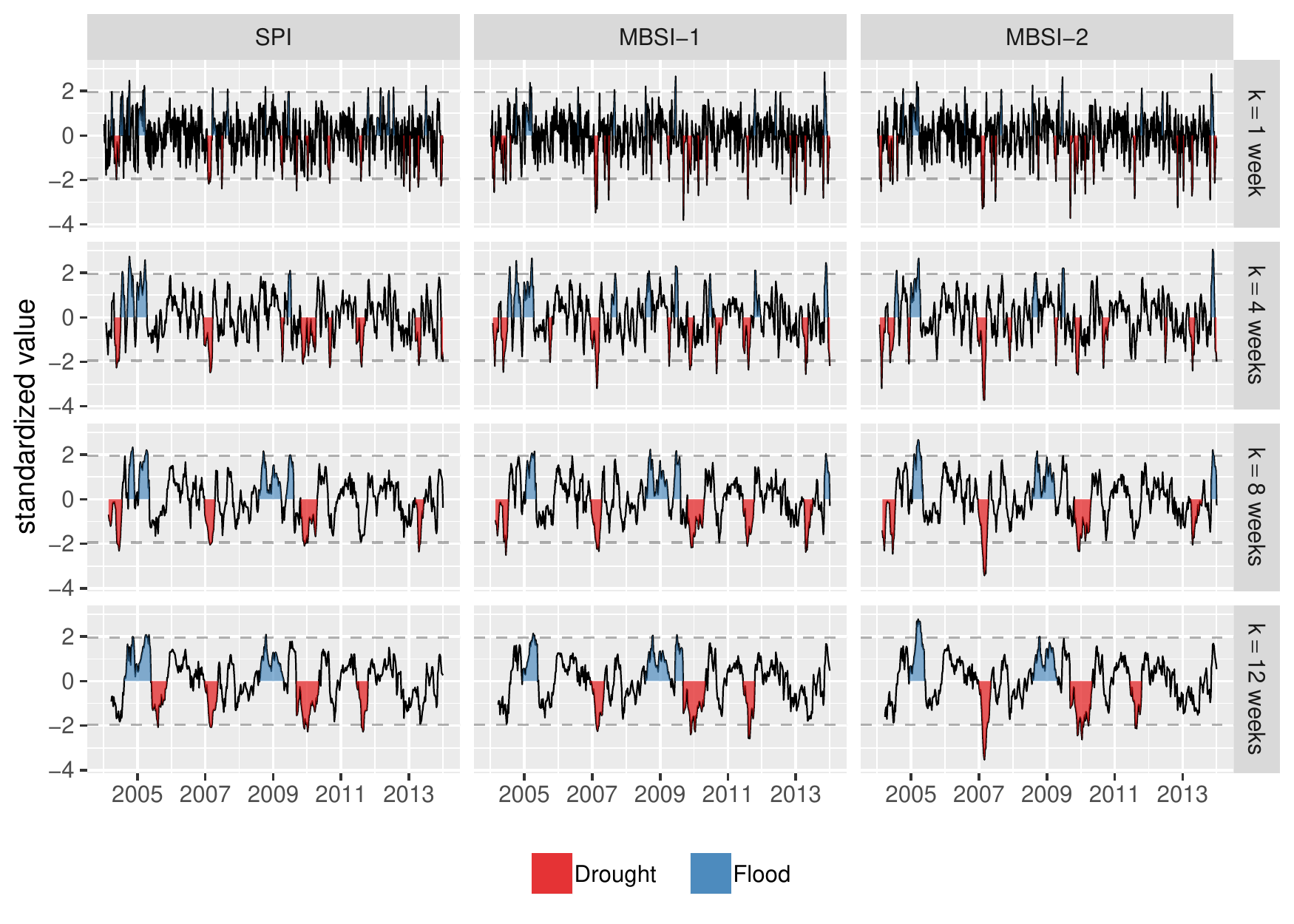}
  \caption{Standardized precipitation values and identification of extreme hydroclimatic events at different time-scales using the SPI and MBSI: the threshold to be considered extreme event was $\pm 1.96$}
  \label{mbsi_fig:spi_extreme_events_mckee}
\end{figure}

The general trend of the standardized precipitation values obtained by three methodologies are similar; however, the actual standardized values corresponding to the identified events, using Section \ref{mbsi_sub:floods_droughts_monitoring}, differs (Figure \ref{mbsi_fig:spi_extreme_events_mckee}). For example, at the time scale of 1 week, most of the identified droughts have, clearly, greater absolute standardized values when working with the \mbsi{}s. We can also see that the number of identified events varies between the methods. For instance, more droughts are identified with the \mbsi{}s when selecting a threshold of $\pm 1.96$ for a time-scale of $8$ weeks. Another difference among the methods is that the \mbsi-2 tends to intensify more the extreme events. For example, it can be seen that, for time-scales of $8$ and $12$ weeks, the levels of the standardized precipitation for $2005$ and $2007$ are more extremes for the \mbsi-2 than the SPI and \mbsi-1.

Amazonas State experienced a well-documented major flood in 2009 and large-scale severe drought in 2010 \citep{Chen2010, Lewis2011}. The two events are highlighted at 8 and 12 weeks time-scales, but they are more emphasized when using the \mbsi-1. For this reason and because it holds properties quite similar to SPI improving the methodology, we preferred to use the \mbsi-1 for further studies on cities of the Brazilian Amazonia. However, we encourage the development of indices like the \mbsi-2 where the model is imposed on the original process under study and analyse another process of interest (such as the moving average process) that depends on the original one, using theoretical properties derived from the initial model. This avoids the need to re-fit the model at different time-scales of potential interest.

\section{Discussion and Conclusions}
\label{mbsi_sec:discussion_spi}

We compared the SPI with two proposed approaches \mbsi-1 and \mbsi-2 to obtain standardized precipitation values. It has been seen that the three approaches are adequate in terms of model assumptions; however, we found some differences that leaded as to select the \mbsi-1 to be used in our studies conducted in the Brazilian Amazonia. Our results clearly demonstrate that the methodology of the SPI can be adapted and placed in a modelling framework that can resolve some of the disadvantages of this index.
\begin{itemize}
  \item Because we use the GAMLSS framework, several distributions can be easily applied to compute standardized precipitation values and the diagnostic of the GAMLSS framework can be used to test model adequacy. Alternatively, it is suggested to evaluate the adequacy of the method by checking the property of the probability integral transform.
  \item The definition of time-scale is more general in the \mbsi-1 and so with this model, it is not necessary to work on the monthly scale. In addition, the observed series of precipitation data (or other quantity of interest e.g. river levels) could have missing values or it might be observed at irregular intervals.
  \item By borrowing strength from temporal autocorrelation and seasonal patterns, the \mbsi-1 can compute standardized precipitation values using a shorter length of records, i.e. less then 30 years, while the SPI usually requires a longer series or a wider time-scale to avoid overfitting.
  \item The \mbsi-1 is a temporally continuous model for precipitation and as such, parameters in the model change more naturally (i.e. smoothly) over time. In addition, the \mbsi-1 could be extended to evaluate extreme events, assume trends over the time, or to incorporate spatial effects.
\end{itemize}

\bibliographystyle{apalike}
\bibliography{library}

\end{document}